\documentclass{aa}
\usepackage{graphicx}
\def\Msol{\thinspace\hbox{$\hbox{M}_{\odot}$}}

\begin{document}
   \title{On the Formation of Massive Stellar Clusters}

%   \subtitle{I. Overviewing the $\kappa$-mechanism}

   \author{Guillermo  Tenorio-Tagle
          \inst{1}
          \and
          Jan Palou\v s
          \inst{2}
          \and
          Sergiy Silich
%          \inst{1}
          \inst{3}
          \and
          Gustavo A. Medina-Tanco
          \inst{4}
          \and
          Casiana Mu\~noz-Tu\~n\'on
          \inst{5}
%          \fnmsep
%          \thanks{Just to show the usage
%          of the elements in the author field}
          }

   \offprints{J. Palou\v s}

   \institute{Instituto Nacional de Astrof\'\i sica Optica y
Electr\'onica, AP 51, 72000 Puebla, M\'exico;
              \email{gtt@inaoep.mx}
         \and
             Astronomical Institute, Academy of Sciences of the Czech
           Republic, Bo\v cn\' \i \ II 1401, 141 31 Praha 4, Czech Republic;
             \email{palous@ig.cas.cz}
         \and
             Instituto Nacional de Astrof\'\i sica Optica y
             Electr\'onica, AP 51, 72000 Puebla, M\'exico;
%            Main Astronomical Observatory National Academy of Sciences of 
%       Ukraine, 03680 Kiyv-127, Golosiiv, Ukraine;
             \email{silich@inaoep.mx}
         \and
            Instituto Astron\^omico e Geof\'{\i}sico, USP, Av. Miguel
St\'efano 4200, 04301-904 Sao Paulo, Brazil;
             \email{gustavo@iagusp.usp.br}
         \and
            Instituto de Astrof\'{\i}sica de Canarias, E 38200 La
Laguna, Tenerife, Spain;
             \email{cmt@ll.iac.es}
%             \thanks{The university of heaven temporarily does not
%                     accept e-mails}
             }

   \date{Received 13 November 2002 / Accepted 14 July 2003}

\abstract{
Here we model a star forming factory in which the continuous creation
of stars results in  a highly concentrated, massive
(globular cluster-like) stellar system. We show that under very
general conditions a large-scale gravitational instability in the 
ISM, which triggers the collapse of a massive cloud, leads with the aid
of a spontaneous first generation of massive stars, to a standing, 
small-radius, cold and dense shell. Eventually, as more of the
collapsing matter is processed and incorporated, the shell becomes 
gravitationally unstable and begins to fragment, allowing the
formation of new stars, while keeping its 
location. This is due to a detailed balance established 
between the ram pressure from the collapsing cloud which, together with 
the gravitational force exerted on the shell by the forming 
cluster, acts against the mechanical energy deposited by the collection 
of new stars. We present a full analysis of feedback and show how the 
standing shell copes with the increasing mechanical energy 
generated by an increasing star-formation rate. The latter 
also leads to a rapidly growing number of ionizing photons, and we
show that these 
manage to ionize only the inner skin of the standing star-forming shell. 
We analyze the mass spectrum of fragments that result from the
continuous fragmentation of the standing shell and show that its shape 
is well approximated at the high mass end by a power law with slope
-2.25, very close to the value that fits the universal IMF. Furthermore, 
it presents a maximum near to one solar mass and a rapid change
towards a much flatter slope for smaller fragments. The
self-contamination resultant from the continuous generation of stars 
is shown to lead to a large metal spread in massive 
($\sim$ 10$^6$ M$_\odot$) clusters, while clusters with a mass similar 
to 10$^5$ M$_\odot$ or smaller, simply reflect the initial metalicity 
of the collapsing cloud. This is in good agreement with the data 
available for globular clusters in the Galaxy. Other observables such as the 
expected IR luminosity and the H$_\alpha$ equivalent width caused by
the forming clusters are also calculated.
}

\maketitle
%
%________________________________________________________________

\section{Introduction}

There is a mode of star formation that leads to massive 
(10$^4$ M$_\odot$ - a few 10$^6$ M$_\odot$), densely concentrated 
collections of stars. These have been named young massive clusters, 
super-star clusters and for the more massive ones the term starburst 
has also been used. The clusters are believed to have evolved from 
an interstellar cloud  mass distribution $N \propto M^{-2}$ 
(Elmegreen \& Efremov 1997) and thus young clusters present a 
similar power law (see also Zhang \& Fall 1999) independent of 
the environment, while old (globular) clusters have a mass distribution 
that falls off towards low masses (Harris \& Pudritz 1994), perhaps due to 
evaporation within a Hubble time (Elmegreen \& Efremov 1997).
As pointed out by Ho (1997) young super-star clusters are 
overwhelmingly luminous concentrations of stars that
present a typical half-light radius of about 3 pc, and a mass that ranges from a
few times 10$^4$ M$_{\odot}$ to 10$^6$ M$_\odot$.
The brightest ones have luminosities up to two orders of magnitude 
higher than R136 in 30 Doradus. Similar super-star cluster
properties have been inferred from HST-STIS observations of AGN 
(Colina et al. 2002), and from radio continuum measurements of 
ultra-compact HII regions not visible in optical images, which
indicates that they are the youngest, densest and most highly obscured star
formation events ever found (Kobulnicky \& Johnson 1999; 
Johnson et al. 2001). The massive concentrations imply a high 
efficiency of star formation which even after long evolutionary times 
permits the tight configuration that characterizes them, despite 
the impact of photo-ionization, winds and supernovae,
believed to efficiently disperse the gas left over from star formation.
Thus the self-gravity that results from the high efficiency of star
formation is what keeps the sources bound together. 
This high efficiency also brings us to believe that
the formation of young clusters should be either
a delayed or a very rapid event, to avoid negative
feedback (Larsen \& Richtler 2000).
The observational evidence now points to such massive units of star 
formation ($\sim 10^4 - 10^6$ M$_{\odot}$) present at the excitation centers 
of blue compact and starburst galaxies such as M82
(de Grijs et al. 2001, O'Connell et al. 1995), and  NGC 253 
(Watson et al. 1996) as well
as in galaxies of different types (see also Larsen \& Richtler 2000
and Larsen 1999). Very similar entities have also been 
found in interacting galaxies. Perhaps the best example of these is the
Antennae with a collection of young star clusters with a 
median effective radius of 4 pc and ages of about 7 Myr 
(Whitmore et al. 1999). These galaxies also show 
other much larger entities (with an outer radius of 450 pc) which 
are not addressed in this paper.
This star-forming activity  in which
masses similar to the total gas content found in galactic giant
molecular
clouds are turned into stars, all in a very small volume ($\sim $ few
pc) implies the rapid accumulation of matter before star 
formation and negative feedback affect the collapsing cloud.
Here we suggest that the feedback from continuously created massive 
stars (M$_*$ $\geq$ 10 M$_\odot$) can sustain a 
fragmenting small radius standing shell, giving origin to
concentrated massive stellar clusters, with a universal IMF. 

Formation of a bound stellar cluster from the supershell
expanding out of the gaseous cloud has been discussed by Brown et
al. (1991, 1995). In their concept  stars are formed when the
supershell has swept out the entire cloud and expanded beyond its
original boundary. However, the physical mechanism 
responsible for the supershell fragmentation and formation of the 
second generation of stars
remains uncertain. Here we give a  thorough discussion of the supershell
gravitational stability and show that the standing shell that forms
due to the balance between the ejected mass ram pressure and the
central cluster gravitational pull may become gravitationally unstable
allowing the continuous formation of the new generation of stars.

Section 2 describes the stellar factory, its self-regulation and the 
physics that lead to its closure, once a massive compact cluster has 
formed. Section 3 deals with the spectrum of fragments (sizes and
numbers) expected from the factory model and compares this result with
observed values of the IMF. Self-contamination is analyzed in Sect. 4, 
where strong predictions on the metallicity of the resultant
clusters are given. Section 5 deals with other observables such as the 
expected IR luminosity and the (H$_\alpha$) equivalent width of the 
resultant clusters. Finally, our conclusions are drawn in Sect. 6.

\section{The star-forming factory}

Our model invokes pressure-bounded, self-gravitating, isothermal
clouds, which may become gravitationally unstable if 
sufficiently compressed (Ebert 1955; Bonner 1956). The gravitational 
instability allows a large cloud ($M_{\rm c} \sim 10^4 - 10^6$ M$_\odot$)
to enter its isothermal ($T_{\rm c} \sim$ 100 K) collapse phase (Larson 1969; 
Bodenheimer \& Steigart 1968; Foster \& Chevalier 1993; 
Elmegreen et al. 2000), thereby developing a density and velocity 
structure with the following characteristics:
1) A central region of constant density (the plateau) where
the velocity increases  linearly from zero km s$^{-1}$ at the
center, to a maximum value ($v_{\rm max}$ = 3.3 $c_{\rm c}$ where
$c_{\rm c}$
$\sim $1 km s$^{-1}$ is the sound speed of the collapsing cloud)
at the boundary. 2) A region of increasing size and constant maximum 
velocity ($v_{\rm max}$), where the density falls off as $R^{-2}$ (the skirt).
As the collapse proceeds and the density in region 1 becomes
larger, the knee region in the density distribution, that separates
zones (1) and (2), moves closer to the center of the
configuration with an increasing speed.  As the density in the plateau
region ($\rho_{\rm p}$) 
increases, unstable fragments begin to form. These will first 
(as $\rho_{\rm p}$ grows larger than 10$^{-20} $ g cm$^{-3}$) have a Jeans 
mass similar to those of massive stars: 
%----------------------------------------------------------------
$
M_{\rm Jeans} (\rm g) = 8.5 \times 10^{22}
\left(\frac{T_{\rm c}}{\mu}\right)^{1.5} \rho_{\rm p}^{-0.5} . \
$
%----------------------------------------------------------------

Although the above equation does not take into account turbulence and 
magnetic fields as in Mac Low \& Klessen (2003), one can, as a first approach,
assume that a first stellar generation with a sufficient number of massive 
stars ($M_*$ =100 - 10 $M_{\odot } $) will form spontaneously in the central plateau region
where the 3D converging flow may trigger their gravitational
instability. The stellar fragments will detach from the flow and, on
time-scales of the order of a few times 10$^5$ yr, will enter the
main sequence. From then onwards, through their winds and terminal 
supernova (SN) explosions, they will begin to have an important 
impact on the collapsing cloud. For this 
to happen however, massive stars ought to form in sufficient numbers as to
jointly stop the infall at least in the most central regions of the plateau.
Otherwise, individual stars, despite their mechanical energy input rate,
will unavoidably be buried by the infalling cloud, delaying the impact of 
feedback until more massive stars form. We thus assume that the first
generation of massive stars is able to regulate itself by displacing and 
storing the high-density matter left over from star formation into a
cool expanding shell, thereby limiting the number of sources in
the first stellar generation. One can show that, given the high
densities ($n \geq 10^4$ cm$^{-3}$) attained both in the wind and
in the plateau region and the size of the latter ($\sim$ 1 pc) a mass $\sim $ 10$^3$ M$_\odot$ is 
available for a first generation of stars (with masses $\geq$ 10 M$_\odot$). 
The shell will be driven by the momentum
injected by the central wind sources, and the ionization front
will rapidly become trapped within the expanding layer (see Sect. 2.1). 
Thus the two most disruptive agents thought to interrupt (Hoyle 1953) or even
stop (Cox 1983; Larson 1987; Franco et al. 1997) the formation of
stars are kept well under control by the infalling cloud.
The large densities also promote a
rapid radiative cooling within the shell, allowing low
temperatures ($\sim$ 10 K). The expanding layer is at all times
confronted with the increasing density and larger
velocity of the matter in the unperturbed plateau region, and it
soon becomes ram-pressure confined. That is, it will soon happen
that the central wind ram-pressure ($\rho_{\rm w} v_{\rm w}^2$) 
exactly balances
the infalling cloud plateau ram-pressure. From then onwards the shell 
of swept-up matter is forced to recede towards the stars, 
given the increasing density and velocity of the undisturbed 
collapsing plateau region,
causing eventually the burial of the first stars. A more interesting
situation arises if the number of sources in the first stellar generation
is such that the final position of the layer of shocked matter 
is close to the knee of the
density distribution (R$_{\rm k}$), where both the infalling gas density 
($\rho_{\rm k}$) and velocity $(v_{\rm max}$) attain their maximum values.
There, near $R_{\rm k}$, the wind ram-pressure ($\rho_{\rm w} v_{\rm
w}^2$) will exactly 
balance the infalling cloud ram-pressure ($\rho_{\rm k} v_{\rm
max}^2$) when the 
mechanical luminosity (L$_{\star}$) of the 
first spontaneous stellar generation
approaches the  critical value
%---------------------------------------------------------------
\begin{equation} \label{eq.2}
L_{\rm crit} =  2 \pi R_{\rm k}^2 \rho_{\rm k} v_{\rm w} v_{\rm max}^2
\end{equation}
%----------------------------------------------------------------
Star formation will suddenly stop as all plateau matter left over from star 
formation is now locked in the standing shell. Once the shell acquires this 
standing location, it will be able to process the infalling cloud mass. In 
this way, the mechanical energy deposited by the first generation of massive 
stars favors the accumulation of infalling cloud mass  
in the standing shell, which becomes gravitationally unstable.

In our steady-state model everything happens at the same time. 
Gravitationally bound fragments continuously form in the unstable 
shell (at $R = R_{\rm k}$) and then, due to their negligible  cross-section, 
freely fall towards the center of the configuration as they evolve 
into stars. The larger number of sources continuously enhances the 
mechanical luminosity of the forming cluster and with it the amount 
of mass returned as a wind into 
the shell (${\dot M_{\rm w}}$). At the same time, the
continuous fragmentation of the shell and the infall of the
resultant fragments acts as a source of mass in the most central
region of the collapsing cloud, and this rapidly modifies the
balance previously established between the wind and the infalling
gas ram pressures. Indeed the ram pressure exerted by the wind
sources, in order to keep the shell at its standing location, will
now have to balance not only the infalling gas ram pressure but also the
gravitational force exerted on the shell by the increasing  mass
of the  central star cluster:
%---------------------------------------------------------------
\begin{equation} \label{eq.5}
4 \pi R_{\rm k}^2 \rho_{\rm w} v_{\rm w}^2 = 4 \pi R_{\rm k}^2 
\rho_{\rm k} v_{\rm max}^2 +
\frac{G M_{\rm sh} M_{\rm sc}}{R_{\rm k}^2}.
\end{equation}
%----------------------------------------------------------------

In the steady-state regime  considered here, the central star
cluster mass, $M_{\rm sc}$, and the shell mass, $M_{\rm sh}$, are
$M_{\rm sc}
= 4 \pi R_{\rm k}^2 \rho_{\rm k} v_{\rm max} t$ and $M_{\rm sh} = 4
\pi R_{\rm k}^2 \Sigma_{\rm sh}$,where $\Sigma_{\rm sh}$ is the shell surface
density and $t$ is the evolutionary time.
Comparing the first and the second terms on the right-hand side of
Eq. (\ref{eq.5}) one can show that very soon, after $t
\ge v_{\rm max}/(4 \pi G \Sigma_{\rm sh}) \approx 10^4 - 10^5 $yr, the
infall ram pressure becomes negligible compared to the
gravitational pull provided by the forming cluster. Thus the shell
becomes gravitationally bound with  its mechanical equilibrium
simply given  by the equation
%---------------------------------------------------------------
\begin{equation} \label{eq.8}
4 \pi R_{\rm k}^2 \rho_{\rm w} v_{\rm w}^2 = \frac{G M_{\rm sh} M_{\rm
sc}}{R_{\rm k}^2}.
\end{equation}
%----------------------------------------------------------------
One can show that despite the increasing effective gravity, the
shell remains stable against Rayleigh - Taylor modes, because the
density of the shocked wind is larger than that of the shocked
infalling cloud. Also, nonlinear thin shell instabilities (see
Vishniac 1994) and their induced pressure perturbations will be
overcome by gravitational forces. 

From the new equilibrium condition (Eq. \ref{eq.8}), one 
can find how the ejected mass density grows with time at the standing 
radius $r = R_{\rm k}$
%---------------------------------------------------------------
\begin{equation} \label{a.1}
\rho_{\rm w}(t) = \frac{G M_{\rm sh} \rho_{\rm k} v_{\rm max}}
{R_{\rm k}^2 v_{\rm w}^2} t =
            \frac{4 \pi G \Sigma_{\rm sh} \rho_{\rm k} v_{\rm
max}}{v_{\rm w}^2} t ,
\end{equation}
%----------------------------------------------------------------
and thus determine the mechanical energy input rate, 
$L_{\rm eq}$ = $\frac{1}{2} {\dot M}_{\rm w}(t) v_{\rm w}^2 $, 
required to support 
the shell in its equilibrium state:  
%---------------------------------------------------------------
\begin{equation} \label{eq.10}
L_{\rm eq} = \frac{1}{2} {\dot M}_{\rm w}(t) v_{\rm w}^2 = 
         8 \pi^2 G \Sigma_{\rm sh} \rho_{\rm k} R_k^2 v_{\rm w} v_{\rm
         max} t.
\end{equation}
%----------------------------------------------------------------
i.e., to support the shell against the gravitational pull
exerted by the forming central star cluster, the mechanical
luminosity would have to grow linearly with time. 

A second constraint on the mechanical luminosity arises 
from a consideration of the star formation rate (SFR). This is defined
by the sum of the two sources of mass at the shell:
the rate at which the collapsing cloud is processed by the shell 
(${\dot M_{\rm in}}$), which is a constant, and that rate at which mass is 
ejected by the star cluster (${\dot M_{\rm w}}$), 
which increases linearly with time. Thus,
%---------------------------------------------------------------
\begin{equation} \label{a.2}
SFR(t) = {\dot M_{\rm in}} + {\dot M_{\rm w}} =
         4 \pi R_{\rm k}^2 \rho_{\rm k} v_{\rm max} \left(1 + 
         \frac{4 \pi G  \Sigma_{\rm sh}}{v_{\rm w} t} \right) ,
\end{equation}
%----------------------------------------------------------------
is a function that increases also linearly with time. Such star
formation rate defines the energy deposition ($L_{\rm sc}$), expected from 
the growing central star cluster. Comparing 
$L_{\rm eq}$ (Eq. \ref{eq.10}) with $L_{\rm sc}$, one can then find 
the value of the shell surface density $\Sigma_{\rm sh}$. 
$L_{\rm sc}$ is to be derived from starburst synthesis models 
(Leitherer \& Heckman 1995, Mas-Hesse \& Kunth 1991, Silich et
al. 2002) taking into consideration the SFR prescribed by (Eq. \ref{a.2}) 
and an assumed metallicity ISM of the host galaxy.

Figure 1 shows that the energy input rate derived independently 
from the starburst synthesis models is in reasonable agreement with 
the equilibrium value (Eq. \ref{eq.10}) over a considerable span of time, 
particularly when  $\Sigma_{sh} \approx 0.7 - 1.1$ g cm$^{-2}$. In such
cases both mechanical energy input rates agree to within less than a 
factor of two - three over almost 25-30 Myr.
%----------------------------------------------------------- S_vib
   \begin{figure}
   \centering
   \vglue 0.1cm
   \includegraphics[width=8cm]{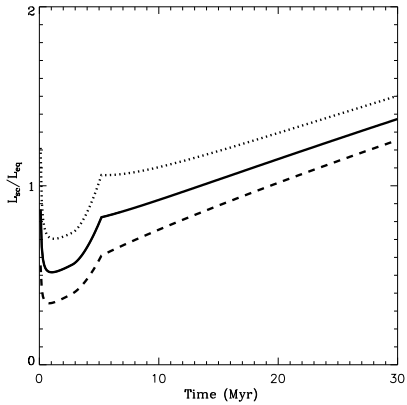}
   \vglue  0.0cm
   \caption{Mechanical energy requirements. The figure shows the ratio 
            $L_{\rm sc}$ (Eq. 6) over  
            $L_{\rm eq}$ (Eq. 5).  
            The $L_{\rm sc}$ values  
            result from starburst synthesis models that assume a SFR(t) 
            as given by Eq. 6, for different values of 
            $\Sigma_{\rm sh}$ = 0.5 (dotted lines), 0.7 (solid lines) and 
            1.1 g cm$^{-2}$ (dashed lines). All models also assume
            upper and lower mass limits equal to 100 M$_\odot$ and 
            1 M$_\odot$ and a slope of -2.25 for the high mass end 
            (as derived in Sect. 3). $L_{\rm sc} $ assumes
            mechanical energy
            input rate from a continuous star formation rate 
            (Leitherer \& Heckman 1995)  
            assuming a galaxy metallicity value = 0.25 Z$_\odot$.}
         \label{fig1}
   \end{figure}
%
%______________________________________________________________

A star-forming factory then results  from a profound
self-regulation that accounts for the mass continuously added to
the forming cluster, as well as for  the mechanical energy that
results from this further addition of mass and its transformation
into stars. Self-regulation keeps the shell at its standing
location and thus with the same fragmenting properties, while the
forming cluster remains hidden behind the shell and the collapsing cloud.

\subsection{The negative feedback caused by photoionization}

The increasing star formation rate also leads to a
rapidly growing number of ionizing photons ($N_{\rm sc}$). 
This has also been 
calculated (see Fig. 2a) using the starburst synthesis model 
(Silich et al. 2002) under the assumption of a linearly increasing 
star formation rate, as prescribed by relation (\ref{a.2}).
Clearly, the ionizing radiation may upset the shell fragmenting properties 
by simply changing, through photoionization, the temperature of the 
swept-up gas. Following Comeron (1997) and Tenorio-Tagle et al. (1999) 
we have derived the impact that such an increasing ionizing photon flux has 
on the collapsing shell.

The shell density has been calculated from the momentum balance:
%---------------------------------------------------------------
\begin{equation} \label{a.3}
\rho_{\rm sh} = \rho_{\rm w}(R_k)\left(\frac{v_{\rm w}}{c_{\rm sh}}\right)^2 .
\end{equation}
%----------------------------------------------------------------
where $\rho_{\rm w}$ as given by Eq. (\ref{a.1}), is a function of
time, and  $c_{\rm sh}$ is the sound speed of the shell. The full 
thickness of the shell is
%---------------------------------------------------------------
\begin{equation} \label{a.4}
l_{\rm sh} = \frac{\Sigma_{\rm sh}}{\mu_{\rm n} n_{\rm sh}} ,
\end{equation}
%----------------------------------------------------------------
where $\mu_{\rm n}$ and n$_{\rm sh}$ are the mean mass per particle 
and the shell 
number density, respectively. Thus the number of photons
required for a complete ionization of the shell is:
%---------------------------------------------------------------
\begin{equation} \label{a.5}
N_{\rm crit} = \frac{4 \pi}{3} \left[(R_{\rm k} + l_{\rm sh})^3 -
           R_{\rm k}^3\right]
           n_{\rm sh}^2 \alpha_{\beta} ,
\end{equation}
%----------------------------------------------------------------
where $\alpha_{\beta} = 2.59 \times 10^{-13}$ cm$^3$ s$^{-1}$ is
the recombination coefficient to all levels but the ground state
(Osterbrock, 1989). A comparison of the critical number of photons 
required for a complete ionization of the shell, with that
of ionizing photons ($N_{\rm sc}$) emitted by the growing central star
cluster ($N_{\rm sc}/N_{\rm crit}$) as function of evolutionary time
($t$) is shown in Fig. 2b. Throughout the evolution, this always
remains $\ll 1$. This implies is  that the ionization 
front, despite the increasing number of photons, 
is trapped within the shell structure and furthermore,  is only 
able to photo-ionize a narrow inner section, the inner skin, 
of the gravitationally unstable shell. It is worth noticing that a
fraction of the UV flux will be absorbed within a free wind region,
reducing further  the number of the UV photons reaching the shell per 
unit area and per unit time. Therefore the two possible negative 
feedback mechanisms, the ionizing radiation and the mechanical 
energy of the forming cluster, remain at all times under control by 
the star-forming factory.
%----------------------------------------------------------- S_vib
   \begin{figure}
   \centering
   \vglue 0.1cm
   \includegraphics[width=8cm]{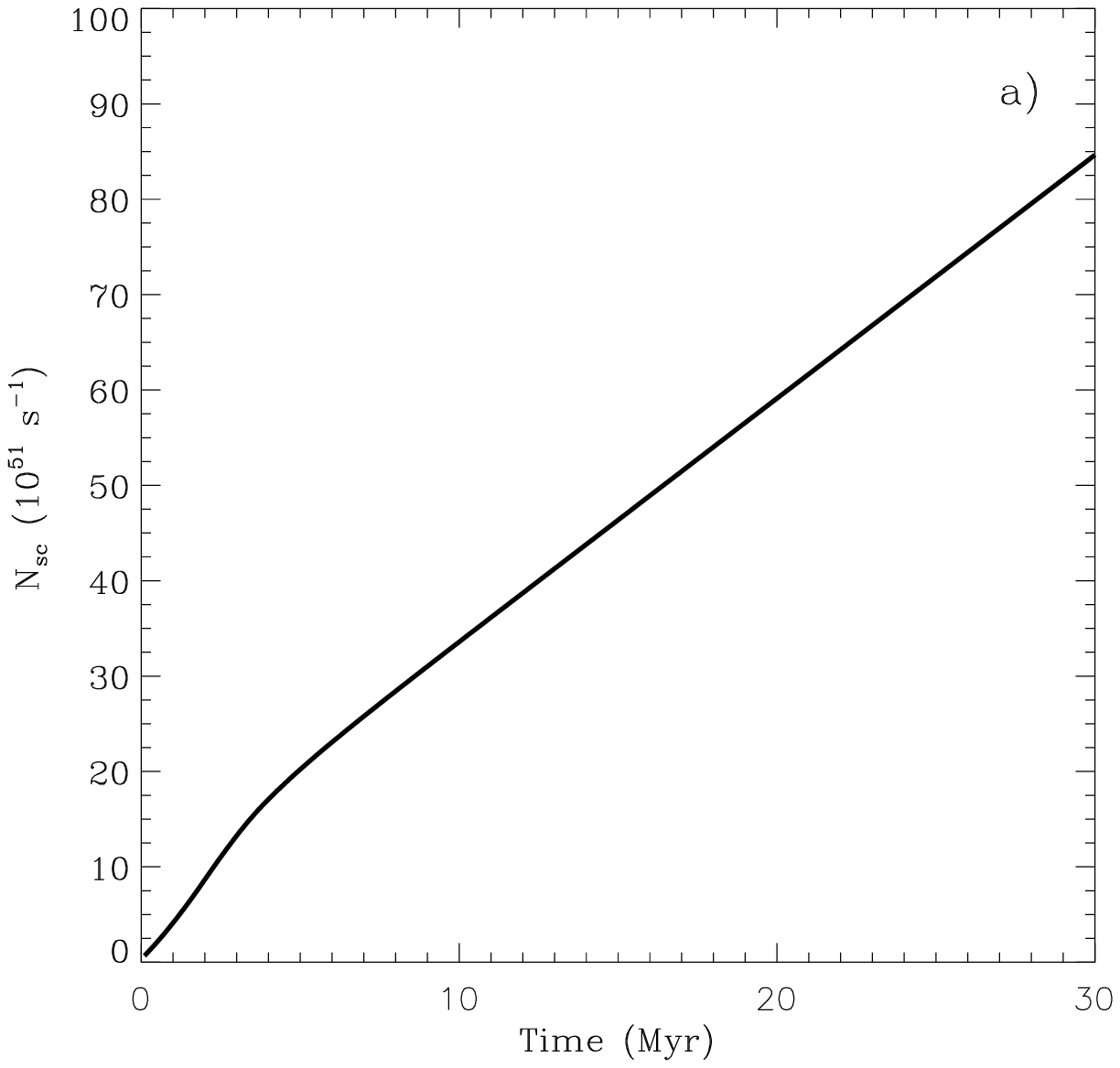}
   \includegraphics[width=8cm]{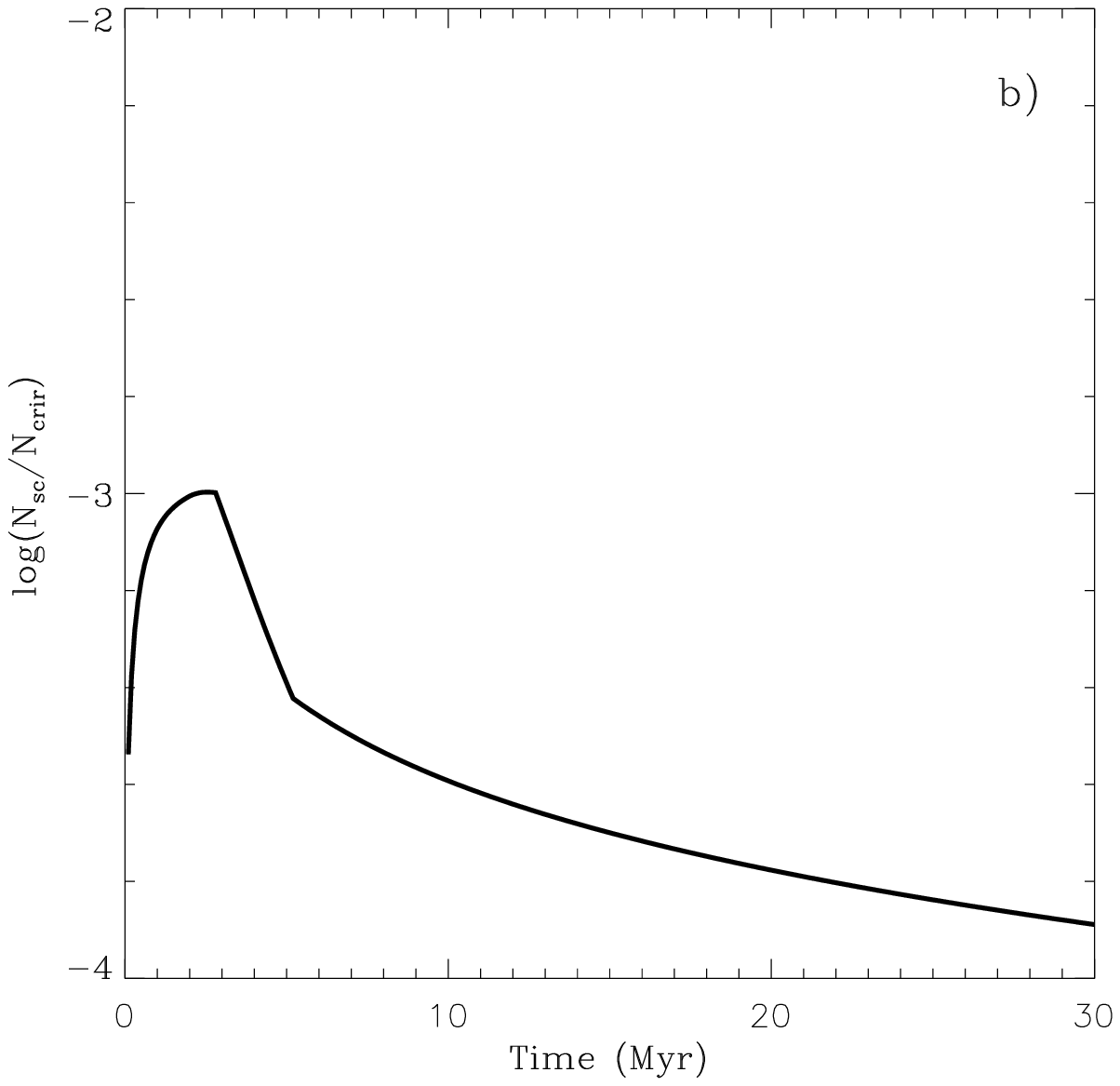}
   \vglue  0.0cm
   \caption{The effects of photoionization. a) The ionizing radiation 
   produced by the forming cluster ($N_{\rm sc}$) as a function of time as 
   derived from the starburst synthesis models under the same
   assumptions as those used in Fig. 1 and a 
   $\Sigma_{\rm sh} = 0.7$ g cm$^{-2}$; b) a comparison of the
   ratio of the photon flux ($N_{\rm sc}$) and the number required to 
   fully ionized the star forming shell ($N_{\rm crit}$), as a 
   function of time.}
         \label{fig2}
   \end{figure}
%
%______________________________________________________________

\subsection{The size of the resultant clusters}

The factory stops operating either because  small clouds ($M_{\rm c}$
$\leq$ $10^5$ M$_\odot$) are rapidly processed by the standing
shell or, in the case of larger clouds, because  the mechanical
energy input rate implied by the SFR condition ($L_{\rm sc}$), here 
derived using starburst synthesis models, begins to largely exceed 
the luminosity required ($L_{\rm eq}$) to keep the shell in its 
standing location. 
This latter possibility arises after 25 Myr  of evolution 
(see Fig. 1), when the luminosity generated by the increasing
SFR begins to dominate the equilibrium condition, although 
not by more than a factor of 2. After this time the shell will lose 
its standing location and be disrupted as it accelerates into the 
skirt of the remaining cloud. Thus, the size of the largest resultant 
clusters in our factory model, is restricted  to a few 
10$^6 $M$_\odot$, the amount of cloud mass that can be processed by the 
standing shell within this time interval.

Consequently, the half-light radius of the resultant cluster
corresponds to the radius of the standing shell, and to the fraction of
mass of the remaining cloud which is accelerated and removed from the
original cloud when the mechanical energy input rate $L_{\rm sc}$
surpasses the equilibrium condition. Assuming that 1/2 of the original
cluster mass has been removed, the radius of the cluster doubles,
reaching the value of a few pc. The final size of the cluster
depends on its subsequent internal evolution and on the environment in the
home galaxy.

Let us assume a cloud of 10$^6$ M$_\odot$ that has developed the
plateau-knee-skirt structure during its isothermal ($T_c = 100$ K)
collapse phase. By the time the knee reaches a radius of 2 pc,
$\rho_{\rm p}$ would be $\sim 10^{-20}$ g cm$^{-3}$, and massive stars
(M$_*$ $\leq$ 100 M$_\odot$) will begin to appear at the center of
the collapsing configuration. These will store the surrounding gas
into an expanding shell, limiting the number of sources in the
first stellar generation. The expanding shell becomes ram-pressure
confined within the plateau. However, given the increasing density
in the unperturbed plateau region, the shell must recede until it
finds the knee of the density distribution, where both density and
velocity of the incoming gas remain at a constant value. Here we
assume a final shell standing at a distance of 1 pc. The density
$\rho_k$ would then be $\approx 6.68 \times 10^{-20}$ g cm$^{-3}$,
and the infalling velocity equals 3.9 km s$^{-1}$.
At this time the outer radius of the 10$^6$ M$_{\odot}$ cloud is
R$_{\rm max} \approx M_{\rm c}/(4 \pi \rho_{\rm k} R_{\rm k}^2) 
\approx 90$ pc. Assuming
$v_{\rm w}$ = 10$^8$ cm s$^{-1}$, the mechanical luminosity derived from
Eq. (1) is $L \approx 5.8 \times 10^{37}$ erg s$^{-1}$. 

In the steady-state regime, the infalling mass ends up being
transformed into stars at the rate at which it is processed by the
standing shell
${\dot M}_{\rm in} = 4 \pi R_{\rm k}^2 \rho_{\rm k} v_{\rm max} 
\approx 4.4 \times
10^{-2}$ M$_{\odot}$ yr$^{-1}$. An originally 10$^6$ M$_\odot$
cloud would then be processed in about 24 Myr, while the mechanical
luminosity of the forming cluster increases almost linearly 
with time up to 6.3$\times 10^{40}$ erg s$^{-1}$. 

Given the self-similar solution of the isothermal collapse phase,
all collapsing clouds are processed in a very similar manner, all with 
a similar IMF, and all of them causing a similar SFR rapidly increasing 
with time. Thus the only 
limitation on the mass of the resultant 
clusters is set by the time during which they are able to enhance 
their mechanical luminosity with time, to keep the fragmenting shell 
at its standing location. Note also 
that for a given  plateau density, collapsing clouds present
the same size plateau, regardless of the mass of the collapsing
cloud ($M_{\rm c}$). Thus, as the radius of the star-forming shell 
is independent of the collapsing cloud mass, the mass-radius relation observed for 
GMCs is not reflected in the stellar clusters, as pointed out by Ashman 
\& Zepf (2001).

\section{The mass spectrum of fragments}

%                                     Two column figure (place early!)
%______________________________________________ Gamma_1 (lg rho, lg e)
   \begin{figure}
   \centering
   \vglue -0.2cm
   \includegraphics[width=8cm]{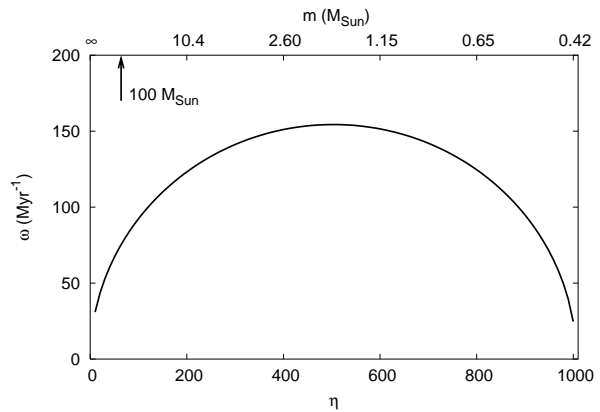}
   \vglue -0.2cm
   \caption{The dispersion relation $\omega (\eta )$ for $c_{\rm sh}$ 
corresponding to 
$T_{\rm sh}$ = 10 K, $R_{\rm k}$ = 1 pc and $\Sigma_{\rm sh}$ = 
0.7 g cm$^{-2}$. 
$\eta _{\rm max} = 503$, corresponds to the wavelength $\lambda $ = 0.01 pc, 
gravitationally unstable mass 
$m = \pi \Sigma_{\rm sh} \lambda ^2 = 1.6 M_{\odot }$, and the 
e-folding time of the growth of perturbations of $5 \times 10^3$ yr. The
largest and smallest masses form at $\eta = 62$ and 1005
corresponding to the mass range 0.39 - 150 M$_{\odot}$. The upper
axis indicates the (non-linear) mass scale, and the arrow the
position of a 100 M$_{\odot}$ fragment.}
              \label{fig3}%
     \end{figure}
%---------------------------------------------------------------

Mass accumulation leads to the gravitational instability of the
standing shell with a well-defined mass and number of resultant
fragments. The dispersion relation  for gravitational instability 
of an expanding shell of radius $R$ is (Elmegreen 1994; W\"unsch 
\& Palou\v s 2001): 
%----------------------------------------------------------------
$
\omega = -{3 \dot R \over R} + \left ({\dot R^2 \over R^2} - {\eta
^2 c^2_{\rm sh} \over R^2} + {2 \pi G \Sigma_{\rm sh} \eta \over
R}\right)^{1/2}, \
$
%----------------------------------------------------------------
where $\Sigma_{\rm sh}$ is the unperturbed surface density
of the shell, $c_{\rm sh}$ its sound speed and $G$ is the
gravitational constant. The condition for instability demands
$\omega$ to be real and positive. The wavenumber $\eta$ is related
to the perturbation wavelength $\lambda $ by $\eta = {2 \pi R
\lambda^{-1}}$, and the e-folding time of the perturbation growth
is ${\omega^{-1}}$. For a standing shell configuration (with $\dot R = 0$) 
this reduces to
%---------------------------------------------------------------------
\begin{equation} \label{eq.44}
\omega ^2 = -{\eta^2 c^2_{\rm sh} \over R^2} + {2 \pi G \Sigma_{\rm sh}
\eta \over R}.
\end{equation}
%---------------------------------------------------------------------

The dispersion relation $\omega(\eta)$ is shown in Fig.
3. The fastest growing mode occurs for $\eta_{\rm max} = {\pi G
\Sigma_{\rm sh} R \over c^2_{\rm sh}}$, corresponding to $\omega_{\rm max} =
{\pi G \Sigma_{\rm sh} \over c_{\rm sh}}$.
Fragments of mass
%---------------------------------------------------------------------
\begin{equation}\label{mass}
m = \pi \Sigma _{\rm sh} \lambda ^2 = 4 
\pi^3 R^2 \Sigma_{\rm sh} \eta^{-2}. \
\end{equation}
%---------------------------------------------------------------------
\noindent form with a frequency proportional to the growth rate,
which is given by $\omega$, and is proportional to the shell
surface available to accommodate fragments of a given wavelength,
${R^2 \over \lambda^2}={\eta^2 \over 4 \pi^2}$. Consequently the
number of fragments $\Delta N$ formed per unit time, out of the 
standing shell, with the wavenumber from the interval 
$(\eta, \eta + \Delta \eta)$ is
%---------------------------------------------------------------------
\begin{eqnarray}
\label{N} \nonumber
& & \hspace{-0.3cm}
\Delta N = \left(\omega \frac{2 \eta}{4 \pi^2} + \frac{\eta^2}{4 \pi^2}
\frac{{\rm d}\omega } {{\rm d}\eta }\right) \Delta{\eta }  = 
      \\[0.2cm]
      & & \hspace{0.5cm}
{\eta \over 4 \pi ^2} \left(2 \omega + \eta \frac{{\rm d}\omega}
{{\rm d}\eta}\right)
\Delta{\eta}
\end{eqnarray}
%---------------------------------------------------------------------
The $\Delta\eta$ interval, can be expressed in terms of the
corresponding $\Delta m$ by means of Eq. (\ref{mass}). And
thus, Eq. (\ref{mass}) and (\ref{N}) imply the mass spectrum
of fragments, or define the initial mass function  $\xi$ (m) =
$\Delta N / \Delta m$.
%---------------------------------------------------------------------
\begin{eqnarray}
      \nonumber
      & & \hspace{-0.3cm}
\xi (m) = Q \pi^{9/4} \Sigma_{\rm sh}^{3/2} R^2 m^{-9/4} \times
      \\[0.2cm]
      & & \hspace{0.5cm}
\frac{-3 \pi^{1/2} c_{\rm sh}^2 m^{-1/2} + 2.5 G \Sigma_{\rm sh}^{1/2}}
{(-\pi^{1/2} c_{\rm sh}^2 m^{-1/2} + G \Sigma_{\rm sh}^{1/2})^{1/2}} ,
\end{eqnarray}
%---------------------------------------------------------------------
\noindent where Q is the normalization factor fixed by the star formation 
rate (SFR). Note that a negative $\Delta m$ should be used for positive
$\Delta \eta$.
Thus, mass accumulation leads to the gravitational instability of
the standing shell with a well defined mass and number of
resultant fragments. From the dispersion relation of the
linearized analysis of the hydrodynamical equations on the surface
of the standing shell, the mass spectrum of gravitationally bound
fragments presents a slope equal to -2.25 for massive objects. The
distribution flattens in the neighborhood of $\eta_{max}$ and
peaks at $m = {\pi c_{\rm sh}^4 \over 4 G^2 \Sigma_{\rm sh}}$.
The minimum mass, obtained from the condition that $\omega$ is positive,
lies at $2 \eta_{\rm max}$. These results are in good agreement with the stellar 
mass distribution (see Fig. 4) inferred for star clusters 
(Moffat 1997; Hunter et al. 1997; Wyse 1997; Piotto \& Zoccali 1999; Paresce \& De Marchi 2000)
and for the solar neighborhood (Salpeter 1955; Scalo 1986; Binney 
\& Merrifield 1998; Kroupa 2001, 2002).

%
%                                                One column figure
%----------------------------------------------------------- S_vib
   \begin{figure}
   \centering
   \vglue -1.0cm
   \includegraphics[width=8cm]{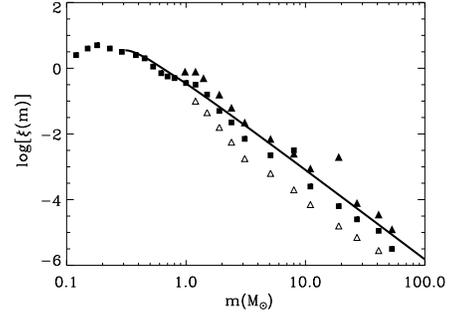}
   \vglue -0.7cm
      \caption{The IMF - $\xi$ (m) - as given by Binney \& Merrifield
   (1998), is compared 
with our results from Eq. (13) normalized to the total cluster mass 
(solid line). The comparison assumes a 10$^6$ M$_\odot$ cloud, used below 
as an example, fully  processed into stars  during a time span of 24 Myr
(see text)}
         \label{fig4}
   \end{figure}
%
%______________________________________________________________

\section{Self-contamination}

After $\sim 3$ Myr of evolution, the mass ejected from the star
cluster (${\dot M}_{\rm w}$) becomes rapidly contaminated by the
supernova explosions. We assume that on reaching the shell the 
metal-enriched matter is immediately mixed with the gas coming from the 
collapsing cloud (${\dot M}_{in}$) which, depending on the host 
galaxy, may present a low metallicity. This leads to a continuously 
increasing abundance for every new generation of stars resulting from 
the fragmenting shell.

We assume that the gas ejected by stellar winds and SNe includes all
the newly synthesized metals as well as metals distributed throughout
the stellar hydrogen envelopes of the progenitor. The  total
mass ejection rate ${\dot M}_{\rm w}$ and the rate of metal ejection 
${\dot M}_{met}$  for a central star cluster with a power-law initial 
mass function then are
%--------------------------------------------------------------------
\begin{eqnarray}
      \label{a.6a}
      & & \hspace{-0.3cm}
{\dot M}_{\rm w}(t) = \sum_i \frac{M_i}{\Delta t}
\frac{M(t+\Delta t)^{2-\alpha} - M(t)^{2-\alpha}}
{M_{\rm low}^{2-\alpha} - M_{\rm up}^{2-\alpha}} ,
      \\[0.2cm]  \nonumber
      & & \hspace{-0.3cm}
{\dot M}_{\rm met} = 
\frac{(\alpha -2)}{M_{\rm low}^{2-\alpha} - M_{\rm up}^{2-\alpha}} \times
      \\[0.2cm]  \label{a.6b} 
      & & \hspace{-0.3cm}
\sum_{\rm i} \frac{M_{\rm i}}{\Delta t}
\int_{M(t+\Delta t)}^{M(t)} Y_{\rm met}(m) m^{-\alpha} {\rm d}m ,
\end{eqnarray}
%--------------------------------------------------------------------
where $M_i$ is the mass of every new generation of stars, $t$ is the
evolutionary time, $\Delta t$ is the time step, $M_{\rm low}$ and
$M_{\rm up}$
are the lower and upper cut-off masses, respectively, $M(t)$ is the
mass of the stars exploding after an evolutionary time $t$ (see Silich
et al. 2002). 

The evolution of the iron abundance of the forming star cluster 
(which can be 
compared with the available metallicities of galactic super-star 
clusters) can be approximated by means of an iron yield $Y_{Fe}$. 
Here we  use the Thielemann et al. (1992) model and extrapolate their 
results as  constant 
yields for low ($< 13$ \Msol) and high ($> 25$\Msol) mass stars (see for 
details Silich et al. 2001).

We then adopt a helium normal abundance (one helium atom for every ten
hydrogen atoms) and calculate the number of hydrogen and iron atoms
mixing inside a shell at every time step
%--------------------------------------------------------------------
\begin{eqnarray}
      \label{a.7a}
      & & \hspace{-0.3cm}
{\dot N}_{\rm H} = \frac{{\dot M}_{\rm tot}}{1.1 \mu_{rm n}} ,
      \\[0.2cm]  \label{a.7b} 
      & & \hspace{-0.3cm}
{\dot N}_{\rm Fe} = \frac{{\dot M}_{\rm Fe}}{A_{\rm Fe} m_{\rm H}} ,
\end{eqnarray}
%--------------------------------------------------------------------
where $\mu_{\rm n} = \frac{14}{11} m_{\rm H}$ is the mean mass per particle, 
$A_{\rm Fe} = 55.4$ is the atomic number of iron in hydrogen mass units,
and $m_{\rm H}$ is the hydrogen mass. The total mass input rate 
${\dot M}_{\rm tot}$, as well as the iron mass input rate ${\dot
M}_{\rm Fe}$, 
includes both the matter ejected as a function of time by the star 
cluster and the mass coming from the collapsing cloud:
%--------------------------------------------------------------------
\begin{eqnarray}
      \label{a.8a}
      & & \hspace{-0.3cm}
{\dot M}_{\rm tot}(t) = {\dot M}_{\rm w} + {\dot M}_{\rm in} ,
      \\[0.2cm]  \label{a.8b} 
      & & \hspace{-0.3cm}
{\dot M}_{\rm Fe}(t) = {\dot M}_{\rm Fe,w} + {\dot M}_{\rm Fe,in} .
\end{eqnarray}
%-------------------------------------------------------------------- 
The iron abundance of every new generation of stars is
%---------------------------------------------------------------------
\begin{equation}
\label{a.9}
\left[\frac{Fe}{H}\right] = log\left(\frac{{\dot N}_{\rm Fe}}
     {{\dot N}_{\rm H}}\right) - log\left(\frac{N_{\rm Fe}}
     {N_{\rm H}}\right)_{\odot } ,
\end{equation}
%---------------------------------------------------------------------
where the Solar iron abundance is $log\left(\frac{N_{\rm Fe}}
{N_{\rm H}}\right)_{\odot } +12 = 7.448$ (Holweger, 2001). 
Fig. 5 presents the
calculated iron abundance of the fragmenting  shell, as a function of
time, for
galaxies with different ISM metallicities. As the clusters form from
the ISM gas as that is contaminated by the supernova products 
from former stellar generations, the resultant metal spread in 
a cluster depends strongly on its final mass (upper axis in Fig. 5). 
In this way, clusters with a final mass smaller than 
2.7 $\times$ 10$^5$ M$_\odot$, processed within 3 Myr, will display the
metal abundance that would  reflect that of the ISM 
at the moment 
of formation. More massive clusters however, as their formation time  
may exceed the supernova era ($\sim$ 3 Myr) from former stellar 
generations, will display a large metal spread in their sources. 
A cluster of 10$^6$ M$_\odot$ (indicated by an arrow in Fig. 5), 
requires 24 Myr to complete its formation. If such a cluster  
forms in a low metallicity (Z$_{\rm ISM} = 0.01$Z$_{\odot}$) galaxy for 
example, it will present  stars with different [Fe/H] abundance
within the range $-2 \le [Fe/H] \le -0.13$, while equally massive 
clusters forming out of a more metal-rich ISM will show a 
correspondingly smaller relative metallicity spread (see Figure 5).
Note however that the spread caused during formation of the clusters 
depends strongly on the assumed metal yields.
%----------------------------------------------------------- S_vib
   \begin{figure}
   \centering
   \vglue 0.1cm
   \includegraphics[width=8cm]{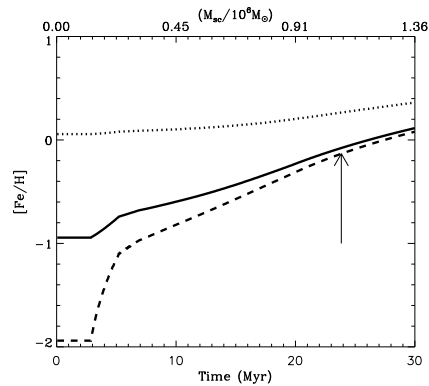}
   \vglue  0.0cm
   \caption{The [Fe/H] metallicity of the resultant clusters.
   The star forming shell metallicity as function of time
           for galaxies with different initial metal abundance: 
           $Z_{\rm ISM}=0.01Z_{\odot}$ (dashed line)
           $Z_{\rm ISM}=0.1Z_{\odot}$ (Solid line) and
           $Z_{\rm ISM}=Z_{\odot}$ (dotted line).
           The resultant clusters, depending on their mass (upper
           axis), will show a metal spread
           that would strongly depend on the ISM original metallicity. 
           }
         \label{fig5}
   \end{figure}
%
%______________________________________________________________

\section{Further observational properties}

Since the star-forming shell is dense, let us assume that there is 
enough dust to obscure the optical emission of the forming star
cluster and 
transform it into the IR. The expected IR nebular luminosity would 
then result from the stellar ionizing flux and from the degraded  
cluster mechanical luminosity, thermalized at the reverse shock  
and then radiated away via effective gas cooling. Therefore the 
luminosity of the IR nebula associated  for example to 
a $\sim 10^6 $\Msol \, cluster may reach $\sim 10^8$L$_{\odot}$.
%(see Figures 1 and 2)

After formation (see Sect. 2.2) the newly formed star cluster
will become visible in the optical line emission region.
At such a  time, the  H$_\alpha$ luminosities of a $10^5$\Msol \, 
and $10^6$\Msol \, star clusters would be 
L$_{\rm H\alpha} \approx 1.4 \times 10^{40}$ erg s$^{-1}$ and  
L$_{\rm H\alpha} \approx 9.4 \times 10^{40}$ erg s$^{-1}$, respectively, 
if all ionizing photons are trapped within the surrounding gas.
The Lyman continuum rate of the $10^6$\Msol \, cluster is around
$6.9 \times 10^{52}$ s$^{-1}$, comparable to the  Lyman continuum
rate ($3 \times 10^{52}$ s$^{-1}$) derived from the radio emission 
flux of super-nebula NGC 5253 (Gorjian et al. 2001).

The continuous creation of stars in the star-forming factory model 
also leaves a foot-print on its H$_\alpha$ equivalent width. Figure 6a
displays the predicted H$_{\alpha}$ equivalent width 
 as a function of 
the star cluster mass, at the moment at which formation is completed
and the cluster becomes visible for the first time ($W_0[\rm H_{\alpha}]$).
The continuous creation of stars during the formation phase
leads to low-mass clusters with initial H$_\alpha$
equivalent widths ($W_0[\rm H_{\alpha}]$) larger than those arising from 
more massive clusters (see Fig. 6a).
 
After formation the number of UV photons drops rapidly and the H$_\alpha$
equivalent width accordingly decreases (Fig. 6b). The evolutionary 
plot implies an HII region life-time which is slightly shorter than 10 Myr 
after cluster formation, with an initially smaller H$_\alpha$ equivalent 
width for  larger masses of the resultant star clusters. 

%----------------------------------------------------------- S_vib
   \begin{figure}
   \centering
   \vglue 0.1cm
   \includegraphics[width=8cm]{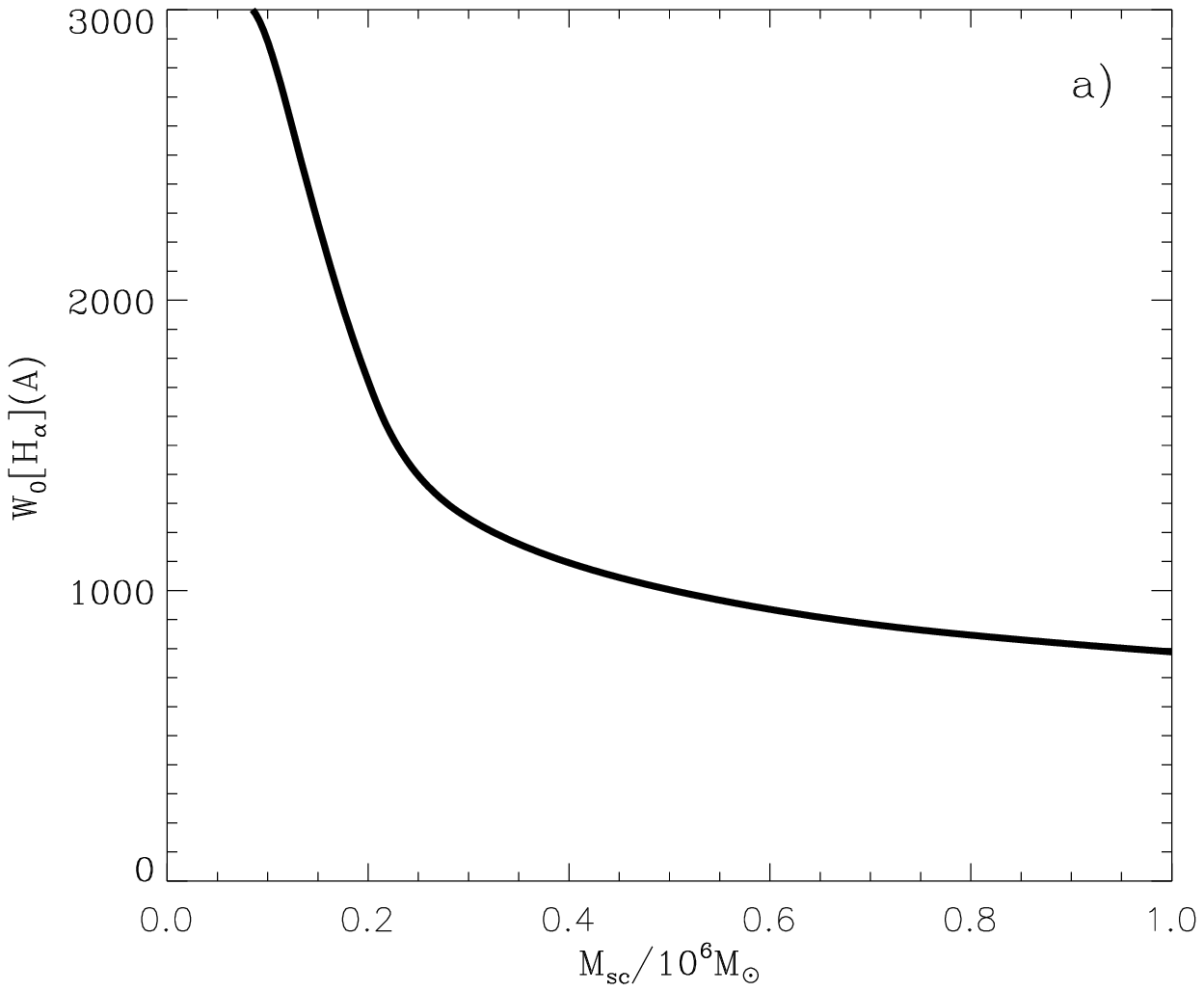}
   \includegraphics[width=8cm]{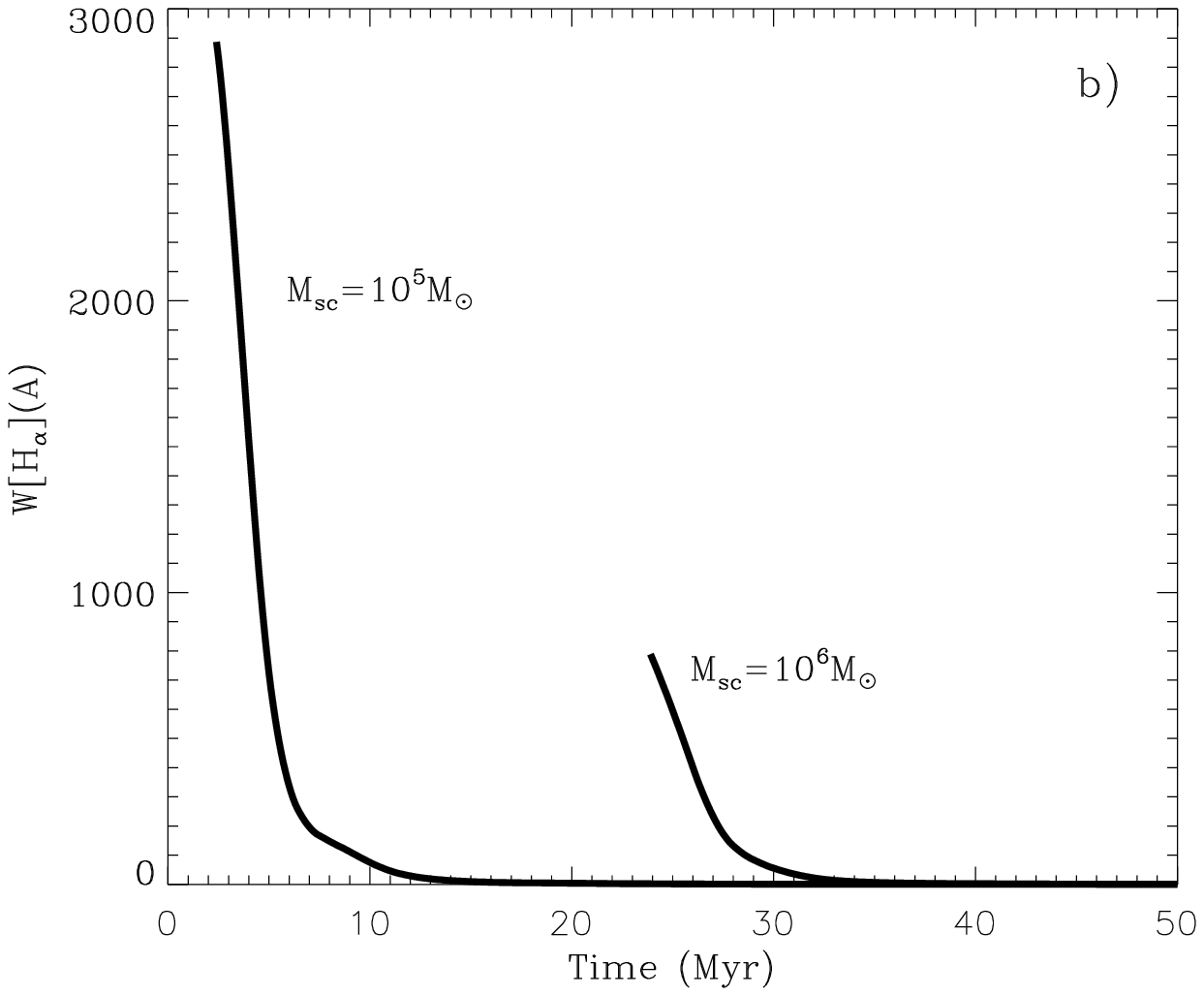}
   \vglue  0.0cm
   \caption{The H$_{\alpha}$ equivalent width. a) The H$_{\alpha}$ equivalent 
   width $W_0$ predicted at the moment when formation is completed and
   the clusters becomes visible, as function of the exciting star
   cluster mass. The synthesis models assumed the same upper and lower 
   stellar mass limits, as well as the increasing star
   formation rate and the slope ($\alpha$ = -2.25) as in Fig. 1.   
   b) The post-formation time evolution of the H$_{\alpha}$ equivalent width  
   for star clusters with a mass of 10$^5$ M$_\odot$ and 10$^6$ M$_\odot$.}
         \label{fig6}
   \end{figure}
%
%______________________________________________________________  

\section{Conclusions}

The formation of compact and massive stellar clusters  
is naturally explained in the framework of the
star-forming factory.  The shell that stores the collapsing cloud is
able to maintain its standing 
location and fragmenting properties, thanks to the larger energy input rate
that results from the increasing number of massive stars in the
central region. With this energy input rate, the forming cluster
is able to balance the increasing gravitational pull  exerted on
the standing shell by the continuously growing mass in stars within the
central region. Self-regulation in the factory can only be
sustained for up to 20 - 25 Myr, while the mechanical energy input
rate grows to satisfy the required energy to keep the shell in its
standing location. Afterwards,  the energy input rate grows and
exceeds the equilibrium condition, causing the acceleration of the 
shell and with it its disruption as it moves into the skirt of the 
collapsing cloud.  After that time, a few times 10$^6$ M$_\odot$ would
have been converted into stars, all following a similar IMF. Smaller 
proto-cluster clouds are fully processed on shorter time-scales. 

Small clusters ($\leq$ 2.7 $\times 10^5$ M$_\odot$) would present the
metallicity identical to the metallicity of the 
ISM out of which they formed. More massive clusters however, will
carry the footprint of self contamination, produced by the supernovae 
products from former stellar generations used to support the factory at work.
Massive clusters will then present a metallicity spread that would
range from their host galaxy ISM metallicity at the moment of formation, to 
values similar to solar metallicity (see Fig. 5).

Upon formation, the more massive the resultant cluster, the smaller 
its initial H$_\alpha$ equivalent width. This may differ by factors of three 
between clusters with a final stellar mass of 10$^5$ and 10$^6$ M$_\odot$.
Large IR luminosities ($\geq$ 10$^8$ L$_\odot$ are predicted for 
factories leading to large ($\geq$ 10$^6$ M$_\odot$) super-star clusters.

A definite prediction of the factory model, applicable in particular
to high mass clusters (say $\sim$ 10$^6$ M$_\odot$), is the
possibility of finding, after formation, a mixture of stellar
populations with different ages and metallicities. For example, within 
a massive cluster the model predicts the co-existence of O stars and 
supergiants evolving at the same time as WR stars, while other stars may 
explode as SN. All of this is the result of the evolution of consecutive 
generations of stars born at different stages during the formation
of the cluster.        

\begin{acknowledgements}
We wish to express our thanks to Caf\'e Slavia and to the Astronomical 
Institute, Academy of Sciences of the Czech Republic (Prague). Also to the
INAOE (Puebla, M\'exico) and the IAG (Sao Paulo, Brazil) for their hospitality
and the atmosphere that favored our collaboration. Our special thanks
to our referee for a report full of ideas and suggestions to improve
our work. Our thanks to B. G. Elmegreen for many clarifying remarks and 
to Monica Rodriguez and Elena Terlevich for their suggestions
regarding the metallicity of our clusters. We also thank D. Rosa for 
his advise regarding the starburst synthesis models and to P. Kroupa
for his comment on the fraction of the gas that can be removed from
the parental cloud. We thank Enrique 
Perez J. and Bill Wall for comments and suggestions. JP gratefully 
acknowledges financial support from the Grant Agency of the Academy of 
Sciences of the Czech Republic under grants No. A3003705, K1048102 and
AVOZ 1003909, 
GMT is partially supported by the Brazilian agencies FAPESP and CNPq, 
GTT, SS  and CMT acknowledge support from the Conacyt (Mexico) grant 
36132-E and the Consejo Superior de Investigaciones Cientificas grant 
AYA2001 - 3939.

\end{acknowledgements}
\vglue -1.0cm

\end{document}